\documentclass[letterpaper,referee]{aa_jd}

\usepackage{graphicx}

\begin{document}

   \title{\hbox{Rotation as a source of asymmetry}\\ in the double-peak
           lightcurves\\
          of the bright EGRET pulsars}

   \author{J. Dyks
          \inst{1}
	   \and           
          B. Rudak\inst{1}\inst{,2}	  
          }

   \offprints{J. Dyks, \email{jinx@ncac.torun.pl}}

   \institute{Copernicus Astronomical Center, Toru{\'n}, Poland
         \and
               Dept. of Astronomy and Astrophysics, Nicholas Copernicus University, Toru{\'n}, Poland
              }

   \date{Received ...; accepted ...}

\abstract{
We investigate the role of rotational effects 
in inducing asymmetry present above $\sim 5$~GeV in the double-peak lightcurves 
of the bright EGRET pulsars:
Vela, Crab, and Geminga. According to Thompson~(\cite{thompson2001}),
the trailing peak dominates over the leading peak above $\sim 5$~GeV
consistently for all three pulsars, even though this is not the case  
over the entire energy range of EGRET,
i.e. above $\sim 100$~MeV.
We present the results of
Monte Carlo simulations of electromagnetic cascades 
in a pulsar magnetosphere
within a single-polar-cap scenario 
with rotationally-induced propagation effects of the order
of $\beta$ (where
$\beta$ is the dimensionless local corotation velocity). 
We find that even in the case
of nearly aligned rotators with spin periods of $P\sim 0.1$~s  
rotation may lead to asymmetric (with respect to the magnetic axis)
magnetic photon absorption which in turn leads to
asymmetric gamma-ray pulse profiles.
The resulting features - softer spectrum of the leading peak and the
dominance of the trailing peak above $\sim 5$ GeV - 
agree qualitatively with the EGRET data of the bright gamma-ray pulsars. 
}

\titlerunning{Asymmetry in the lightcurves of EGRET pulsars}
\authorrunning{J. Dyks \& B. Rudak}

   \maketitle 


\section{Introduction}
Good quality gamma-ray data 
for three pulsars - Vela, Crab, and Geminga - acquired with EGRET 
aboard the CGRO tempts to
analyse the properties of pulsar high-energy radiation as a function 
of photon energy and phase of rotation.
Gamma-ray spectra of pulsed radiation from these sources 
(as well as from three other EGRET pulsars:
B1706-44, B1951+32, and B1055-52) 
extend up to $\la 10$~GeV. 
All three pulsars feature gamma lightcurves characterised 
by two strong peaks separated
by 0.4 to 0.5 in rotational phase.
These double-peak pulses are asymmetrical 
and their profiles change with energy.
Above $\sim 100$~MeV the leading peak (LP) 
is stronger than the trailing peak (TP) in the case of the Vela and the Crab pulsars,
and the opposite is true for Geminga.
However, for all three pulsars
their  leading peaks exhibit lower
energy cutoffs - around $\sim 5$~GeV - than the trailing peaks (TP). 
In other words, the trailing peaks dominate over
the leading peaks above $\sim 5$~GeV (Thompson \cite{thompson2001}).
In the case of the Vela pulsar and Geminga, this effect is accompanied by
the softening of the spectrum of the leading peak 
(Fierro et al.~\cite{fmnt}, Kanbach \cite{kanbach99}).
Potential importance of the double-peak pulse asymmetry in the case
of Vela was already acknowledged -- at the time when the COS-B data
became available -- by
Morini~\cite{morini} who attempted to explain the asymmetry
with a hybrid model, with two different mechanisms responsible
for the formation of the leading and the trailing peak.

High-energy cutoffs in pulsar spectra
are interpreted within polar cap models as due to one-photon absorption 
of gamma-rays in strong magnetic field with subsequent $e^\pm$-pair creation.
A piece of observational support for such an interpretation comes from   
a strong correlation between the inferred `spin-down' magnetic field 
strength and the position of the high-energy cutoff (Baring \& Harding
\cite{bh2000}, Baring \cite{b2001}).
This in turn opens a possibility that the observed asymmetry 
between LP and TP, i.e.
the dominance of LP over TP above $\sim 5$~GeV,
is a direct consequence of propagation effects 
(which eventually lead to stronger magnetic photon absorption 
for photons forming LP than TP)
rather than due to some inherent property of the gamma-ray emission 
region itself.

The aim of this paper is to investigate 
the role of pulsar rotation in a built-up of such asymmetry 
in the double-peak pulse profiles.
We consider purely rotational effects: 
due to presence of rotation-induced electric
field $\vec E$, aberration of photon direction and 
slippage of magnetosphere under the photon's path.
In section 2 we compare them with some other effects which may be responsible
for the asymmetry (like various distortions of the magnetic field structure).
In section 3 we show that the rotation effects
result in an asymmetric pair production rate for the leading and the
trailing part of the magnetosphere
even in the case when the magnetic
field structure and the population of radiating particles  are symmetric
around the magnetic pole. 
In section 4 asymmetric pulse profiles are calculated
as a function of photon energy and then 
the model predictions of the ratio of fluxes in the
leading and trailing peaks
are compared with the inferred 
ratio for Vela at different energy bins. 
In section~5 we address the significance of rotation-driven asymmetry 
across the pulsar parameter space.
Our main results are discussed in Section~6.

\section{Symmetric features of the model}

It has been argued in many studies of radio properties of pulsars
(eg.~Blaskiewicz et al.~\cite{bcw91}, Gangadhara \& Gupta \cite{gg2001})
that a rigidly rotating static-like dipole
can be used as a good approximation of dipolar magnetic field
as long as only
most important rotation effects, of the order of $\beta =
v/c$ (where
$v$ is the local corotation velocity and $c$ is the speed of light), are to be 
considered.
According to order-of-magnitude estimates,
small distortions of the dipolar magnetosphere induced by
rotationally-driven currents can be neglected:
longitudinal currents suspected to flow within the open field line region
cannot modify $\vec B$ by a factor exceeding $\beta^{3/2}$ whereas
toroidal currents due to plasma corotation
change $\vec B$ barely by a value of the order of $\beta^2$.
A more comprehensive discussion of the influence of currents on the
magnetospheric structure 
can be found in Beskin (\cite{beskin}) and references therein.

Below we follow the approximation of a rigidly rotating static-like dipole:
at any instant the magnetic field has the shape of a static dipole 
in the frame which corotates with a neutron star.
Moreover, the magnetosphere is assumed to be filled out everywhere
with the Goldreich-Julian charge density, so that a rotation-induced 
electric field $\vec E = -\vec\beta\times\vec B$
is present in the OF, whereas $\vec E^\prime = 0$
in the frame corotating with the star.
We neglect deviations from this corotational electric field which are present
in the charge-deficient polar gap region -- they do not exceed a factor of $\beta^{3/2}$.

In our Monte Carlo simulations (section 4) development of gamma-ray radiation is based on
the model of Daugherty \& Harding \cite{dh82}, with
primary electrons being 
injected along magnetic field lines at an altitude of a few neutron star radii,
at the magnetic colatitude corresponding to 
the last open magnetic field lines, and uniformly in the magnetic azimuth.
The electrons are assumed to accelerate instantly to the energy 
$E_0 \sim 10^7$ MeV
and subsequently to cool down emitting curvature photons. Some of the photons
induce in turn electromagnetic 
cascades which propagate outwards in a form of a hollow cone beam
(see Dyks \& Rudak \cite{dr2000} for detailed description of directionality aspects
of the casades as well as viewing geometry).

\section{Rotation-driven asymmetry}

\begin{figure}
\centering
\resizebox{\hsize}{!}{\includegraphics{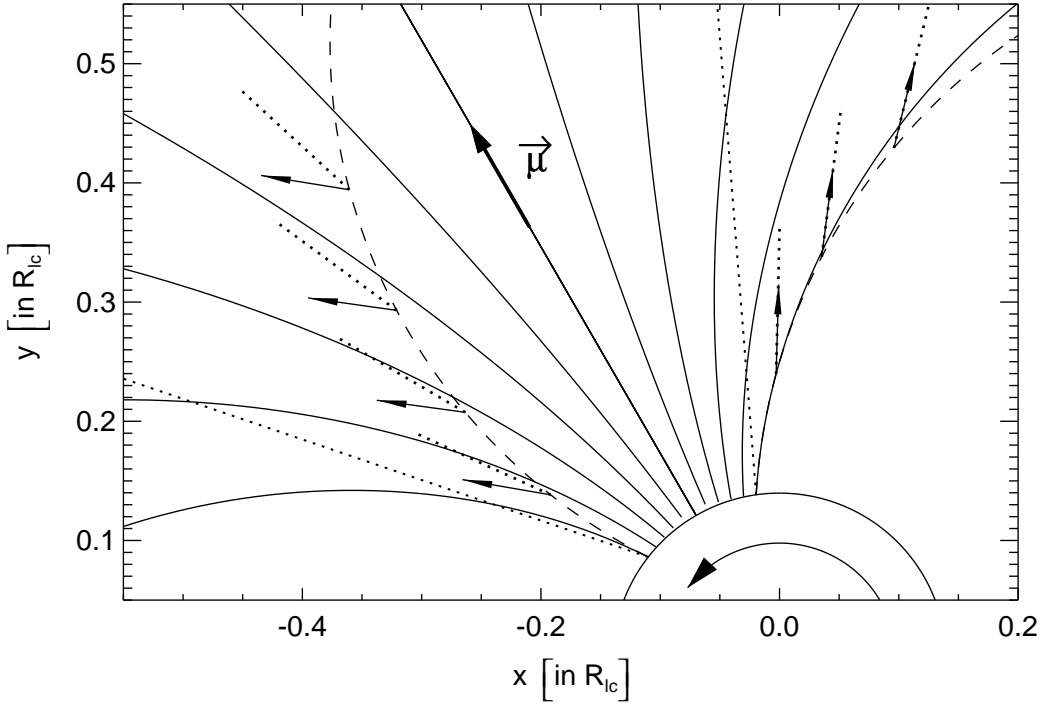}}
\caption{Top view of orthogonally rotating pulsar with the spin period
$P=1.5$ ms; the sense of rotation is counter-clockwise (indicated);
the cartesian coordinates are in units
of the light cylinder $R_{\rm lc} = cP/2\pi$.
The magnetic field lines are approximated with the static-like dipole of 
the magnetic moment $\vec \mu$ (indicated).
Two photon trajectories (starting from two opposite points on the polar cap)
in the inertial observer frame (OF) are marked
with two long dotted lines. These are the prototypes of the leading peak (on the left) and the trailing
peak (on the right).
In the corotating frame (CF)
these trajectories follow the long dashed lines.
Short segments of the dotted-line trajectories (OF) have been transformed via rotation
to several actual photon locations in the
magnetosphere, and thus they 
indicate the local propagation directions $\hat \eta$ of the photons in OF. 
Solid arrows at these locations point along the directions $\hat \eta^{\scriptscriptstyle FP}$ of local free propagation
in OF.
Note much larger angles between $\hat \eta^{\scriptscriptstyle FP}$ and $\hat \eta$ for the photons of the leading
peak than for the photons of the trailing peak.
}
\label{dipole}
\end{figure}

In this section we present in detail the mechanisms which lead to rotation-induced
asymmetry in the (otherwise axially symmetric) hollow-cone gamma-ray beam. 
For the sake of better demonstration of the effects we consider 
an exaggerated case: propagation of a photon in the equatorial 
plane of a fast orthogonal rotator.
Fig.~\ref{dipole} shows the case for the rotation period $P=1.5$ ms:
photons of the same energy in the corotating frame (CF)
are emitted from two opposite points on the outer rim of the polar cap 
tangentially to the magnetic field lines in the CF. 
To follow their straight-line propagation in the inertial observer frame (OF)
three effects have to be taken into account:
1) In the OF the photons propagate at an aberrated direction
(dotted lines in Fig.~\ref{dipole}) and differ in energy.
2) The rotation-induced electric field $\vec E=-\vec \beta \times \vec B$,
which is present in the OF, modifies the rate of the magnetic photon
absorption in a different way for photons forming the LP than
for photons forming the TP.
3) The electromagnetic field in the OF is time-dependent: because of
the rotation the photons propagate through different parts of the
magnetosphere.

\begin{figure}
\centering
\resizebox{\hsize}{!}{\includegraphics{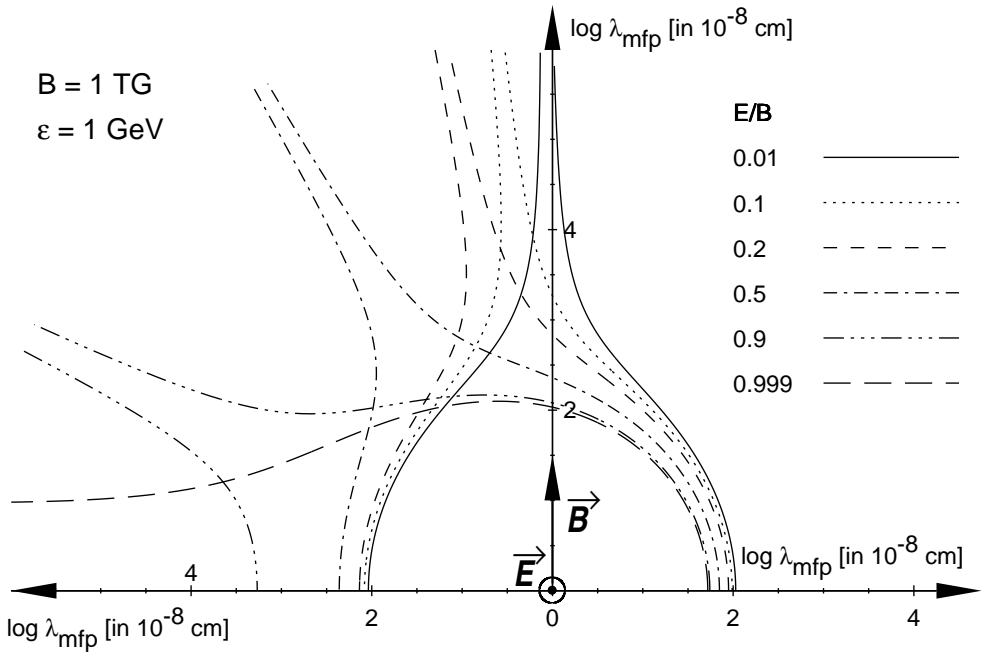}}
\caption{Directional dependence of the mean free path $\lambda_{\rm mfp}$ for magnetic 
absorption of a photon emitted at the center of the reference frame (with the coordinates $[0,0]$)
is shown for six values of $E/B$ (denoted with six different line types).
The electric field vector $\vec E$ is pointing
at the right angle to the page, and toward the reader; the $\vec E\times\vec B$ vector is parallel
to the horizontal axis, and pointing to the left. 
The mean free path $\lambda_{\rm mfp}$  is
symmetric with respect to the horizontal axis.}
\label{mfp}
\end{figure}

We find that the second effect - due to rotationally induced $\vec E$ - plays a dominant role
in generating the asymmetry in the magnetic absorption rate $R$ between 
photons of the LP and the TP. An importance of a weak electric field $\vec E \perp \vec B$
for the rate $R$ was for the first time recognized by Daugherty \& Lerche (\cite{dl75}) who
presented also its quantitative treatment.
A consequence of its presence is that
the rate $R$ does not vanish along the direction of local $\vec B$ any more, and  
becomes non-axisymmetric with respect to it as well.
Instead, the rate $R$ vanishes along two new directions which lie in the plane
perpendicular to $\vec E$ and deviate from $\vec B$ by angle
$\sim E/B$ in such a way that 
in a local coordinate frame with $\hat z \parallel \vec B$, 
$\hat y \parallel \vec E$, and $\hat x \parallel \vec E\times\vec B$ the 
``free propagation" direction $\hat \eta^{\scriptscriptstyle FP} = [\eta^{\scriptscriptstyle FP}_x, \eta^{\scriptscriptstyle FP}_y, \eta^{\scriptscriptstyle FP}_z]$
has two solutions: $[E/B, 0, \pm(1 - E^2/B^2)^{1/2}]$.
For definiteness, hereafter we will consider
photons which propagate outwards within the regions 
above the northern magnetic pole (i.e. 
with propagation vectors $\hat \eta$ satisfying $\hat \eta \cdot \vec B > 0$)
which corresponds to the case of
$\eta^{\scriptscriptstyle FP}_z = +(1 - E^2/B^2)^{1/2}$.
Fig.~\ref{mfp} shows the mean free path $\lambda_{\rm mfp} = c/R$ 
for the magnetic photon absorption for different directions 
in the plane perpendicular to $\vec E$. 
The rate $R$ was calculated by performing Lorentz transformation
to a frame in which $\vec E^\prime = 0$ and then applying a purely magnetic
formula. 
The formula of Erber (\cite{er66}) 
with a modification of Daugherty \& Harding (\cite{dh83}) correcting its near-threshold inaccuracy
is used throughout the paper:
\begin{eqnarray}
\lefteqn{R^\prime(\varepsilon^\prime, \vec B^\prime, \sin\theta_B^\prime)
=\null}\nonumber\\
&&\null = c_1
\sin\theta_B^\prime B^\prime \exp{[-c_2 f/ (\varepsilon^\prime \sin\theta_B^\prime 
B^\prime)]},
\label{r0}
\end{eqnarray}
where $\theta_B^\prime = \angle(\hat \eta^\prime, \vec B^\prime)$,
$\varepsilon^\prime$ is the photon energy; $c_1$ and $c_2$ are constant quantities,
while $f = f(\varepsilon^\prime, \vec B^\prime)$ is the near-threshold correction of Daugherty \& Harding \cite{dh83}.
Six lines in Fig.~\ref{mfp} correspond to six values of the ratio $E/B$: $0.01$, $0.1$,
$0.2$, $0.5$, $0.9$, and $0.999$.
Note that the free propagation direction deviates from the local $\vec B$
by angle $\theta^{\scriptscriptstyle FP}$ which increases with increasing contribution of the electric
field: $\theta^{\scriptscriptstyle FP}_z \equiv
\arccos(\eta^{\scriptscriptstyle FP}_z) 
= \arcsin(E/B) \simeq E/B$. 
Moreover, $R$ increases ($\lambda_{\rm mfp}$ decreases) monotonically with
increasing angle $\angle (\hat \eta, \hat \eta^{\scriptscriptstyle FP})$. As long as $E\ll B$, magnetic photon absorption rate $R$
remains approximately symmetric around the free propagation direction $\hat \eta^{\scriptscriptstyle FP}$.       
 
The directions of $\hat \eta^{\scriptscriptstyle FP}$ at various points within the magnetosphere
are shown in Fig.~\ref{dipole} in the OF (with solid arrows).
At the points of photon emission the direction $\hat \eta^{\scriptscriptstyle FP}$
overlaps with the photon direction $\hat \eta$ since  
this is where the angle $\theta_B$ between the photon direction and the magnetic field 
line equals
$\theta_B \simeq E/B$ and consequently
$R = 0$ (Harding et al. \cite{hte78}, Zheng et al. \cite{zzq98}). 
Initially, therefore, the rate $R$
is the same for photons emitted from two opposite sites of the polar cap rim
which in this picture give rise to the leading peak and the trailing peak.
But as the photons propagate outward
their $\hat \eta$ begin to deviate from local
free propagation directions $\hat \eta^{\scriptscriptstyle FP}$ opening thus a possibility for magnetic photon absorption
and electron-positron pair creation. Three effects are responsible for it to happen:  
(1) curvature of the magnetic field lines, (2) increase of $E/B$ with
altitude, and (3) slippage of magnetic field lines under the photon's path.
For a static-like dipole (assumed in section 2)
the curvature of lines is symmetric with respect
to the magnetic axis. Consequently, symmetry is expected between the rates of magnetic absorption
for LP photons and TP photons. 
Any asymmetry between the LP and the TP may occur via the effects (2) and (3) only.
Whenever the curvature of magnetic
field lines happens to be relatively large 
(which is the case at the polar cap rims of pulsars with $P\sim 0.1$ s) 
the effect (1) significantly dominates over (2) and (3), which 
means that the resulting asymmetry is quite subtle.
However, for a relatively small curvature of magnetic
field lines (and appreciable local corotation velocities $\beta$)
the resulting asymmetry becomes pronounced
[note that these 'favourable' conditions are fulfilled within
outer-gap accelerators].

We find that the effect (2) -- the presence of electric
field -- is crucial for generating the asymmetry.
The way in which this effect operates can be most clearly assessed
by inspecting Fig.~\ref{mfp}.
Let us consider a photon propagation vector $\hat \eta$ anchored 
at the origin of frame in the figure.  
For photons emitted  close to the star, both in the LP and the TP, 
$\hat \eta$ initially points along the $\hat \eta^{\scriptscriptstyle FP}$ direction which differs only slightly 
from the direction of $\vec B$ in Fig.~\ref{mfp}
 since $E \ll B$.
As the LP photon moves away from emission point,
its propagation vector rotates clockwise in Fig.~\ref{mfp}
which reflects the fact that $\vec B$ diverges from $\hat \eta$ 
due to the magnetic field line curvature.
At the same time, however, the directional pattern of $\lambda_{\rm mfp}$
rotates counterclockwise due to increase in $E/B$ 
which additionally enhances the absorption rate.
In the case of the TP photon, however, {\em both} 
its propagation vector $\hat \eta$ and the directional pattern 
of $\lambda_{\rm mfp}$ rotate in the same direction
(counterclockwise) with respect to $\vec B$ in Fig.~\ref{mfp}
so that the absorption rate is weakened.
Thus, for photons within the LP the effects (1) and (2)
cummulate, whereas for the TP they effectively tend to cancel out 
each other, and 
the expected asymmetry between the peaks is due to stronger absorption suffered by 
photons within the LP than by photons
of the same energy within the TP.
Accordingly,
the high energy cutoff in the LP spectrum will occur at
a slightly lower energy than the cutoff in the TP spectrum.
The slippage (3) affects this picture in the way which depends on
both the rotation period and the photon position within the magnetosphere,
but an overall picture remains unchanged. For most rotation periods ($P >$ a few ms), 
the slippage reduces the asymmetry only marginally. 
For the fastest rotators ($P \sim 1.5$ ms) it enhances
the asymmetry by making photons of the TP to propagate  
along the free propagation direction (or, equivalently, along the
magnetic field lines in the CF).
The latter case is presented in Fig.~\ref{dipole} where the
photon propagation direction $\hat \eta$ as seen in the OF (dotted lines) and the local 
free propagation direction $\hat \eta^{\scriptscriptstyle FP}$ in the OF (solid arrows) are shown for 
several positions along the photon trajectory in the CF. 
Strong absorption within
the LP is anticipated, whereas within the TP $\hat \eta$ nearly coincides with $\hat \eta^{\scriptscriptstyle FP}$.

Another way to understand this asymmetry is to
follow photon trajectories in a reference frame (with primed quantities) where the condition
\begin{equation}
\vec E^\prime = 0
\label{e0}
\end{equation} 
is fulfilled.  
In such a frame, an asymmetry in $R^\prime$ (cf. eq.(\ref{r0}))
for the LP and TP photons arises from transformation properties of 
$\theta_B^\prime$ (aberration), $\varepsilon^\prime$ (red- or blue-shift),
and $B^\prime$. 
One of the reference frames satisfying the condition (\ref{e0}) is a reference frame of local $\vec E\times\vec B$ drift. 
Denoting dimensionless drift velocity $\vec \beta_{\scriptscriptstyle D} = \vec E\times\vec B/B^2$, 
the transformations read:
\begin{eqnarray}
\varepsilon^\prime = \varepsilon\,\gamma_{\scriptscriptstyle D}(1 -
\eta_x\beta_{\scriptscriptstyle D}), \ \ \ \ \ \ \ \ \ \
\vec B^\prime = \frac{\vec B}{\gamma_{\scriptscriptstyle D}},\nonumber\\
\sin\theta_B^\prime=
\frac{[(\eta_x - \beta_{\scriptscriptstyle D})^2 + \eta_y^2(1 - \beta_{\scriptscriptstyle D}^2)]^{1/2}}{(1 - \eta_x\beta_{\scriptscriptstyle D})}.
\label{t0}
\end{eqnarray} 
In the equatorial plane of orthogonal rotator $\eta_y = 0$
so that $\eta_x = \mp\sin\theta_B$, where the signs `minus' and `plus'
correspond to the leading and the trailing peak, respectively.
The transformation of propagation angle reduces then to 
$\sin\theta_B^\prime = |\eta_x - \beta_{\scriptscriptstyle D}|/(1 - \eta_x\beta_{\scriptscriptstyle D})$.
The Taylor expansion of $\sin\theta_B^\prime$, $\varepsilon^\prime$,
and $B^\prime$ around $\beta_{\scriptscriptstyle D} = 0$ reveals that
a difference between the magnetic absorption rates in the locally drifting frame and in the OF
results primarily from the aberration of photon direction, whereas the changes
in $\varepsilon$ and $\vec B$ are second order effects.
Obviously, the aberration is asymmetric for the LP and the TP ($\eta_x < 0$ and $\eta_x > 0$, respectively).
Fig.~\ref{dipole} presents this `aberration effect' 
in the rigidly corotating frame, where $\vec E^\prime = 0$ is assumed. 
Photon trajectories in this frame
(dashed curves) indicate clearly that photons of the leading
peak encounter larger $B^\prime_\perp$ 
than photons of the trailing peak of the same energy.

\section{Numerical modelling of the gamma-ray data}

We performed Monte Carlo simulations of curvature-radiation-induced electromagnetic cascades
developing
above a polar cap. The cascade development due to magnetic photon absorption accompanied by $e^\pm$-pair creation
and then synchrotron emission was followed in a 3D space in order 
to analyse pulse properties.
As an example we choose a model with basic parameters of the Vela pulsar: 
$B_{\rm pc} = 6.8$ TG, $P=0.0893$ s.
In order to meet observational restrictions for the Vela, both spectral and temporal,
the following general requirements within  
polar-cap scenarios had to be satisfied: 
1) a polar-cap accelerator should be placed a few stellar radii above pulsar's surface (Dyks et al. \cite{alic}); 
2) an inclination angle $\alpha$ of the magnetic dipole with respect to the spin axis
must not be large, and the pulsar has to be a nearly-aligned rotator (Daugherty \& Harding \cite{dh94}).
Recently Harding \& Muslimov (\cite{hm98}) proposed a physical mechanism
for lifting the polar cap accelerator up to $0.5 - 1\ R_{\rm NS}$
above the surface. However, this altitude is still too low 
to explain the $10$ GeV radiation emerging the Vela magnetosphere unattenuated.
Therefore, we placed the polar-cap accelerator at the
altitude of $h_0  = 4\ R_{\rm NS}$ to ensure that the magnetosphere 
is not entirely opaque to curvature photons of
energy $\la 10$ GeV (see Dyks et al. \cite{alic} for the detailed model spectral fitting for the Vela pulsar).
Similarly, Miyazaki \& Takahara (\cite{mt97})
achieved the best agreement between the observed and their 
modelled pulse profiles
of the Crab pulsar
placing the accelerator at $h_0 = 4\ R_{\rm ns}$.
To reproduce
the observed peak-to-peak separation $\Delta^{\rm peak}
\simeq 0.42$ (Kanbach et al.~\cite{kab94})
we assumed (after Dyks \& Rudak \cite{dr2000}) for the inclination angle
$\alpha$ and the observer's angle $\zeta_{\rm obs}$ (an angle between the line-of-sight and
the spin axis) that $\alpha = \zeta_{\rm obs} = 7.6^\circ$.

\begin{figure*}
\resizebox{\textwidth}{!}{\includegraphics{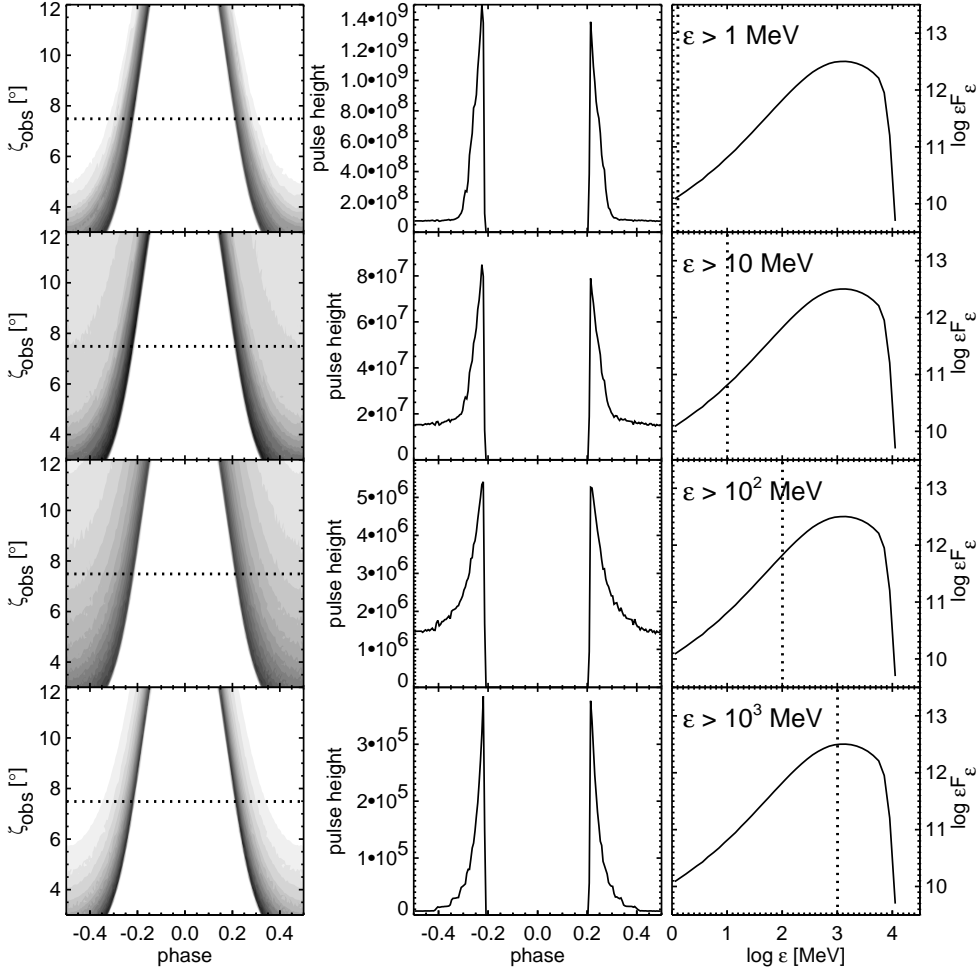}}
\caption{\hskip-2mm {\bf a.} Directional and spectral gamma-ray characteristics 
calculated for the Vela pulsar with the angle $\alpha$ between the spin axis and the magnetic axis
set $\alpha = 7.6^\circ$ (nearly aligned rotator). Eight rows are shown (the continuation in Fig.3b),
with three panels each.
{\bf Left column} shows the outgoing photons of energy $\varepsilon > \varepsilon_{\rm limit}$ 
which are mapped onto the parameter space $\zeta_{\rm obs}$ vs. $\phi$, where
$\zeta_{\rm obs}$ is the viewing angle (between the spin axis and the l.o.s)
and $\phi$ denotes the phase of rotation.
{\bf Middle column} shows the double-peak pulse profile  formed with these photons 
when $\zeta_{\rm obs}=  7.6^\circ$ is chosen (yielding the peak-to-peak separation equal 0.42).
{\bf Right column} shows the phase-averaged 
energy spectrum (the flux level $\varepsilon F_{\varepsilon}$ in arbitrary units) for 
$\zeta_{\rm obs} = 7.6^\circ$ i.e. the same for all rows. 
Dotted vertical line indicates
the part of the spectrum  ($\varepsilon > \varepsilon_{\rm limit}$) which contributes to the 
corresponding pulse profile on the left.
The eight rows correspond to 8 consequtive values of 
$\varepsilon_{\rm limit}$: $1$, $10$, $10^2$, $10^3$, $4\cdot 10^3$, $6\cdot10^3$,
$8\cdot10^3$, and $10^4$ MeV (these values are displayed in the panels of the right column).
}
\label{profiles}
\end{figure*}

Our numerical results are presented in Fig.~\ref{profiles} (a + b). 
The three columns of Fig.~\ref{profiles} 
show (from left to right): 1) mapping onto the parameter space $\zeta_{\rm obs}$ vs. $\phi$ of
outgoing photons with energy $\varepsilon > \varepsilon_{\rm limit}$ (where $\phi$ denotes a phase of rotation),
2) double-peak pulse profile due to these photons when $\zeta_{\rm obs}=  7.6^\circ$, and 3) phase-integrated 
energy spectrum of these photons, with the position of $\varepsilon_{\rm limit}$ indicated
with dotted vertical line.
The eight rows correspond to 8 different values  of 
$\varepsilon_{\rm limit}$: $1$, $10$, $10^2$, $10^3$, $4\cdot 10^3$, $6\cdot10^3$,
$8\cdot10^3$, and $10^4$ MeV (top to bottom).
An asymmetry in the double-peak profiles is noticable even
though the rotator is nearly aligned: at the highest energies, above
$\sim 6$ GeV,  the leading peak LP is less intense than the trailing peak TP
(three lowermost panels in the middle column in Fig.~\ref{profiles}\thinspace b). 
This is a direct result of stronger magnetic absorption
of the LP photons comparing to the TP photons. 
The distribution of these photons in the corresponding panels of 
$\zeta_{\rm obs}$ vs.~$\phi$ (the left column)
shows that at viewing angles $\zeta_{\rm obs}$
larger than $7.6^\circ$ (not allowed due to the fixed peak-to-peak separation of 0.42)
the asymmetry in pulse profile would be even stronger.
This demonstrates an increasing role of rotational effects
as the distance  from the spin axis increases.

\addtocounter{figure}{-1}

\begin{figure*}
\resizebox{\textwidth}{!}{\includegraphics{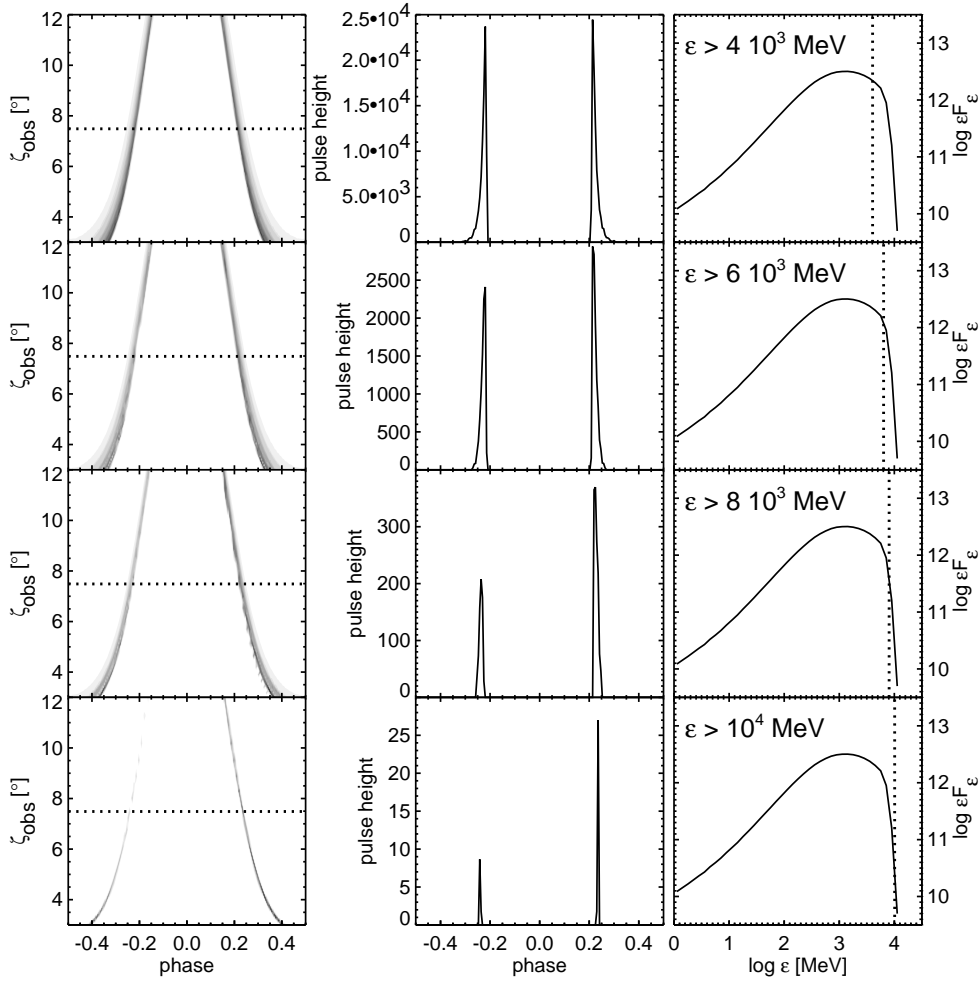}}
\caption{\hskip-2.1mm {\bf b.} Continuation of Fig.~\ref{profiles}\thinspace a.}
\end{figure*}

\begin{figure}
\centering
\resizebox{0.7\textwidth}{!}{\includegraphics{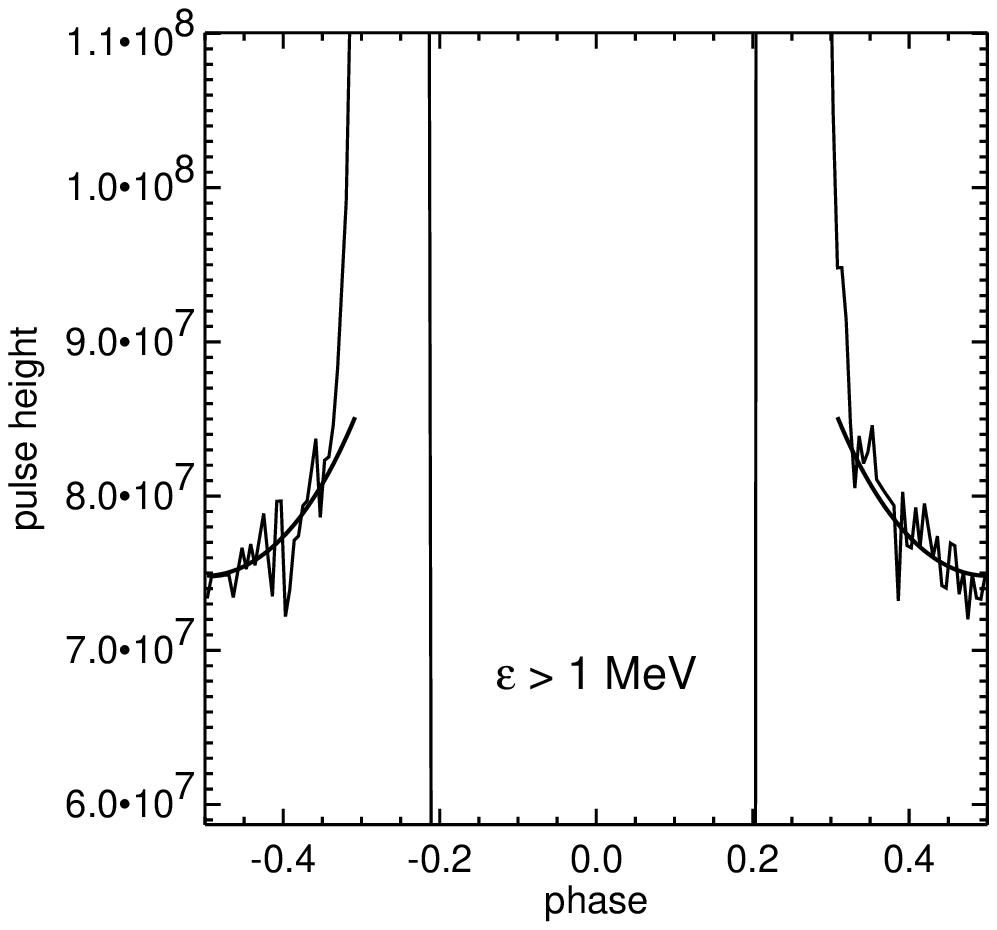}}
\caption{Close-up view of the pulse profile for $\varepsilon > 1$~MeV  from 
Fig.~\ref{profiles}\thinspace a. 
The curvature-radiation wings ($|\phi| > 0.3$) accompanying both peaks can be now
acknowledged.
Their shape is modelled analytically (shown with thick solid lines).}
\label{offpulse}
\end{figure}

In the course of magnetic absorption high-energy curvature photons
are converted into electron-positron pairs which in turn emit low-energy synchrotron photons.
Asymmetry in the absorption rate as
discussed above means, therefore, an identical asymmetry in the $e^\pm$ pair production rate.
Consequently, higher number of low-energy synchrotron photons emerges at the LP than at the TP.
This is the reason for a dominance of the LP over the TP below $\sim 100$ MeV,
noticable in Fig.~\ref{profiles}\thinspace a.
Combining  the results from both energy domains,
a characteristic inversion in the relative strentgh of the LP and the TP occurs across the gamma-ray energy space.
A qualitatively similar inversion of peak intensities takes place in the gamma-ray double-pulse of the Vela pulsar 
(Thompson \cite{thompson2001}).

The beam of synchrotron radiation in our cascades occupies a very narrow range 
of magnetic colatitudes; in other words - it is highly anisotropic. 
The reasons for this include a very limited
range of altitudes at which the $e^\pm$ pairs are created 
and the effects of relativistic beaming.
By comparison, curvature radiation below $\sim 100$ MeV is much less anisotropic.
Therefore, the prominent peaks visible at $\varepsilon < 100$ MeV 
(two uppermost panels of Fig.~\ref{profiles}\thinspace a)
consist almost entirely of synchrotron radiation (SR) photons, whereas the apparently flat wings outside the peaks
(i.e.~within the ``offpulse" region  corresponding to high altitudes) are composed of curvature radiation (CR) photons.
A close-up view of the double-peak pulse profile
for $\varepsilon > 1$ MeV shown in Fig.~\ref{offpulse} reveals that the CR wings  
are not flat  - in fact their intensity decreases with increasing phase $|\phi|$; 
moreover, their shapes can be reproduced  with analytical means:
Spectral power of curvature radiation 
$\frac{{\rm d} P_{\rm cr}}{{\rm d} \varepsilon }$
well below a characteristic photon energy $\varepsilon_{\rm crit}\propto
\frac{\gamma^3}{\rho_{\rm cr}}$ does not depend on the energy of radiating particles $\gamma$ 
but on the curvature radius $\rho_{\rm cr}$ of magnetic field lines solely.
Since primary electrons reside within a pulsar magnetosphere for a limited period of time
$\varepsilon_{\rm crit}$ has a lower limit which equals
roughly $\la 100$ MeV (see Rudak \& Dyks (\cite{rd99}) for details). 
Therefore, the wings in the pulse profiles below $100$ MeV fall off
due exclusively to an increase in the curvature radius $\rho_{\rm cr}$ of magnetic field lines:  
this proceeds according to the following relation
\begin{equation}
\frac{{\rm d} P_{\rm cr}}{{\rm d} 
\varepsilon } \propto \rho_{\rm cr}^{-2/3},
\label{wings}
\end{equation}
which then leads to the smooth solid lines in Fig.~\ref{offpulse}.
As the photon energy increases
and the strength of the synchrotron peaks decreases the curvature wings become more and more pronounced.
They are most noticeable near $100$ MeV. 
Above $100$ MeV  the wings  gradually  disappear (see the middle column in Fig.~\ref{profiles}\thinspace b)
because radiating electrons are not energetic enough at high altitudes.  
At the same time the peaks become narrower  -
an effect noticed by Kanbach et al.~(\cite{kab94}) in the EGRET data for the Vela pulsar.

\begin{figure*}
\centering
\resizebox{\textwidth}{!}{\includegraphics{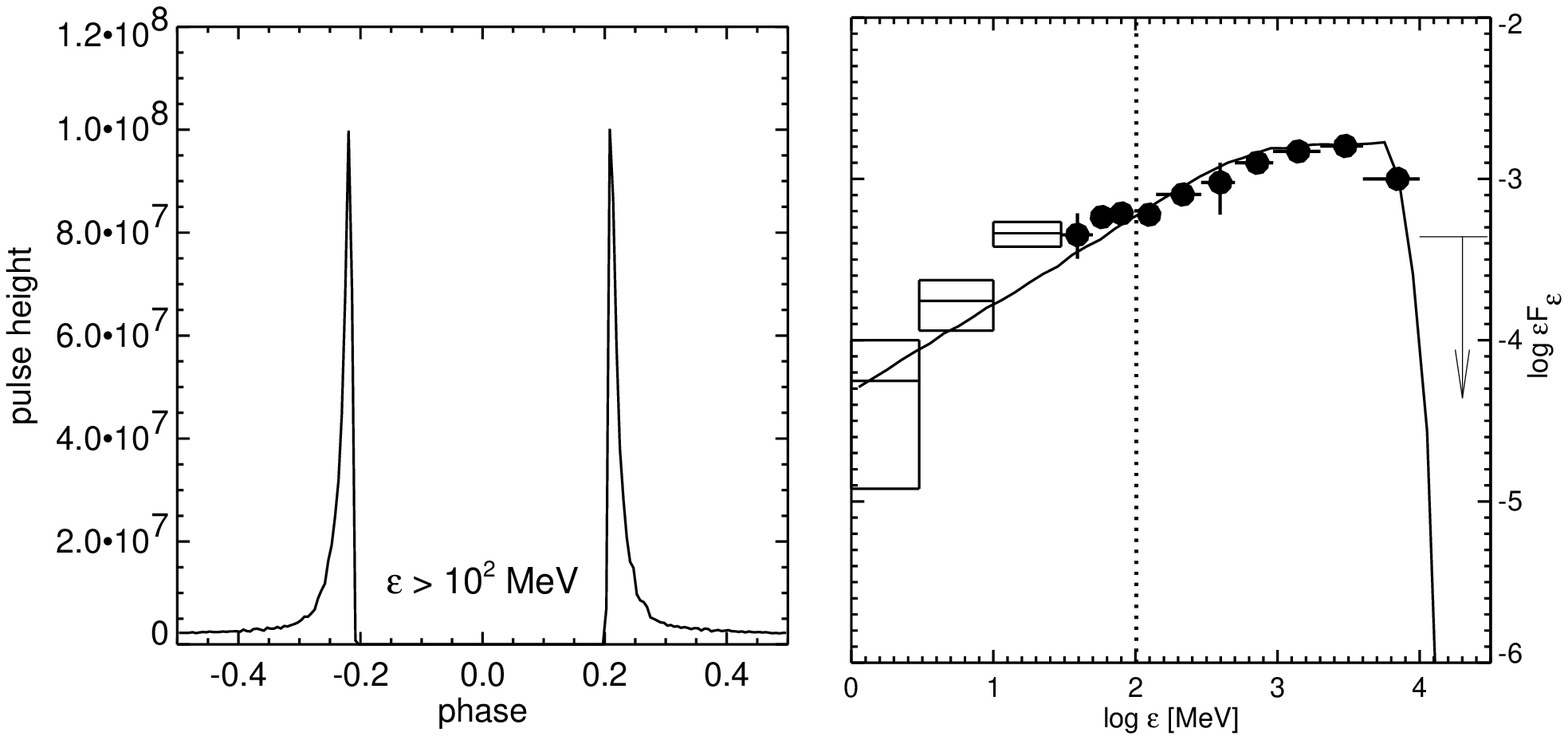}}
\caption{Pulse profile integrated for $\varepsilon > 100$~MeV (left panel) and the  
         phase averaged energy spectrum (right panel) are shown
         for the case of initial energy of primary
         electrons $E_0 = 2\cdot 10^7$ MeV.
         Note the relatively weaker wings outside the peaks
         and much softer spectrum in comparison with the previous case of $E_0 =
         10^7$ MeV (Fig.~\ref{profiles}\thinspace a, third row from the top).
	 The data for the Vela pulsar are  laid over the model spectrum:
         boxes are the data from the first COMPTEL source catalogue
         (see Table 3 in Sch\"onfelder et al.~\cite{schonfelder2000}), and dots  (plus an upper limit
	 above 10~GeV) are 
         the EGRET data (Thompson et al.~\cite{thompson97}). The flux level
	 $\varepsilon F_{\varepsilon}$ is in MeV~cm$^{-2}$~s$^{-1}$ units.}
\label{bigcasc}
\end{figure*}

As noted by Daugherty \& Harding (\cite{dh96}) the wings within the offpulse region must not
be too strong within the entire energy range of EGRET if the theoretical pulse profiles are to
resemble those of the Vela pulsar. 
We find that the intensity of wings relative to the intensity of peaks 
depends sensitively on the richness of the  cascades, i.e. on the multiplicity ${\cal M}$ (the number
of created pairs per primary electron).
The results discussed above and presented in Fig.~\ref{profiles} had been obtained for the initial
energy of primary electrons $E_0 = 10^7$ MeV which 
yielded ${\cal M} = 73$.
By increasing the initial energy $E_0$ up to $2\cdot 10^7$ MeV
the multiplicity reaches ${\cal M} = 830$ and the corresponding pulse profile at $100$ MeV
(left panel of Fig.~\ref{bigcasc})
changes notably with respect to its counterpart of Fig.~\ref{profiles}\thinspace a.
It reveals now a much lower level of wings outside the peaks. 
Equally important is the change in the shape
of the phase-averaged energy spectrum which becomes much softer by gaining more power in the low-energy range  
(right panel in Fig.~\ref{bigcasc}).
Both new features are in rough agreement with the data for the Vela pulsar, contrary to the case with $E_0 =
10^7$ MeV.
It is interesting to note that the association of the broad peaks at 100 MeV
with the relatively hard spectrum (Fig.~\ref{profiles}\thinspace a) 
on one hand, and 
of the narrow peaks with the soft spectrum (Fig.~\ref{bigcasc}) on the other hand
do resemble qualitatively the observed characteristics 
of Geminga and the Vela pulsar, respectively.

\begin{figure}
\centering
\resizebox{0.8\textwidth}{!}{\includegraphics{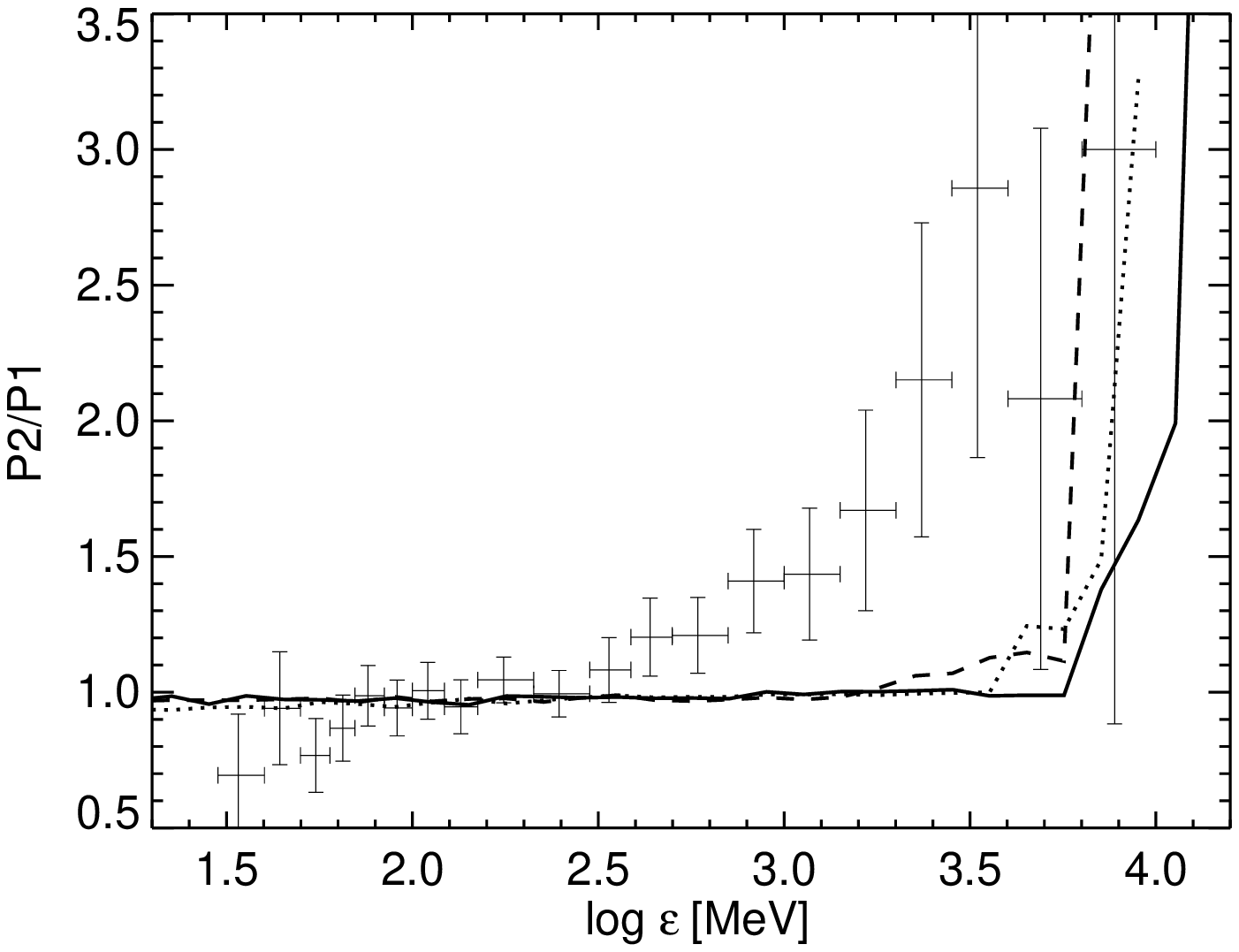}}
\caption{The ratio of the number of photons in the trailing 
peak and the leading
peak $P2/P1$ inferred from the EGRET data for Vela 
is shown in function of photon energy~$\varepsilon$.
The results of model calculations performed for three different values
of the altitude $h_0$ of the accelerator are indicated with three lines:
dashed, dotted, and solid -- for $h_0 = 2 R_{\rm NS}$, $3 R_{\rm NS}$, 
and $4 R_{\rm NS}$, respectively.
}
\label{p1p2}
\end{figure}

We may now test our model of the double-peak asymmetry by
confronting the numerical
results obtained for specific pulsar parameters
with the data for real objects. Since the effect is induced by magnetic
absorption the expected
weakening of the leading peak with respect to the trailing peak
occurs only in the vicinity of the high-energy spectral cutoff.
Therefore, it is essential
to have a good photon statistics also at the highest energy bins, i.e.
above
$1$~GeV. As far as the EGRET data are concerned this requirement is
barely satisfied even for Vela.
With these limitation in mind, we consider Vela as the only appropriate
case to provide the test.
We used the
EGRET data for Vela to calculate the ratio (denoted as $P2/P1$)
of the photon counts
in the LP and the TP (denoted as $P1$ and $P2$, respectively).
For each energy bin (the energy bins
cover the range between $\sim 30$~MeV and $\sim 10$~GeV)   we calculated
$P1$ ($P2$)
by summing up all photons
within the range $\phi_{\rm lp}\pm 0.05$ ($\phi_{\rm tp}\pm 0.05$) in
phase,
where $\phi_{\rm lp}$  ($\phi_{\rm tp}$) is the phase of maximum in the
LP (the TP)
at $100$ MeV.
Fig.~\ref{p1p2}
shows the observational points as well as their estimated
errors\footnote{Detailed analysis of the data
will be presented elsewhere (Wo{\' z}na et al.~2002, in preparation).}
along with
the results of model calculations performed for
three different altitudes: $h_0 = 2 R_{\rm NS}$, $3 R_{\rm NS}$, and $4
R_{\rm NS}$.
The overall qualitative and quantitative  behaviour of $P2/P1$ for the
EGRET data
is very similar to the dependence presented by Kanbach et
al.~\cite{kanbach80} for the COS-B data.
The data points certainly can acommodate our model.
However, to answer the question of whether it would be inevitable to
invoke any additional processes to reproduce
the increase in $P2/P1$ inferred from the data
requires better photon statistics at the spectral cut-off and careful
statistical analysis.

\section{Rotational asymmetry as a function of pulsar parameters}

The asymmetry effects are marginal for nearly aligned pulsars with periods
$P \sim 0.1$ s. For the above-described model of Vela, they are noticeable
only because of the high altitude of the accelerator ($h_0 = 4 R_{\rm NS}$).
However, for highly inclined ($\alpha \ga 45^\circ$) and fast pulsars
($P \la 10$ ms) the rotational effects result in asymmetry of 
considerable magnitude.
If detected by GLAST, GeV emission from such objects would provide 
powerful diagnostics of the polar cap model. Below we present the magnitude of
such asymmetry predicted for a wide range of parameters for fast pulsars.

As a measure of rotational effects we consider the escape energy
$\varepsilon_{\rm esc}$ which is defined as a
maximum energy of a photon (and the photon is emitted tangentially to its `parent' 
magnetic field line; a footpoint of this line has  magnetic colatitude $\theta$) 
for which the magnetosphere is still
transparent, i.e.~for which the optical depth integrated along the photon
trajectory is less than $1$.
This energy was calculated with our numerical code and the results are shown below.
For the sake of comparison with the case of no rotational effects we recall 
a simplified, yet quite accurate\footnote{For 
presentation of detailed numerical results see e.g. Harding et al.~\cite{hbg1997}.} for magnetic 
fields weaker than $\sim 0.1 B_{\rm cr}$, analytic formula which (after e.g. Bulik et al.~\cite{bulik}) reads
\begin{equation}
\varepsilon_{\rm esc}(\theta) = \frac{\theta_{\rm pc}}{\theta} \cdot \varepsilon_{\rm esc}(\theta_{\rm pc}),
\label{eesc0}
\end{equation}
where
\begin{eqnarray}
\varepsilon_{\rm esc}(\theta_{\rm pc}) \approx 1\ {\rm GeV}\ \left(\frac{B_{\rm pc}}
{0.1\,B_{\rm cr}}\right)^{-1}
\left(\frac{\theta_{\rm pc}}{0.01\,{\rm rad}}\right)^{-1}\nonumber\\
\times \left(\frac{r_{\rm m}}{R_{\rm NS}}\right)^{5/2}.
\label{eesc}
\end{eqnarray}
Here $r_{\rm m}$ is the radial coordinate of the emission point
(with spherical coordinates ($r_{\rm m}$, $\theta_{\rm m}$, $\phi_{\rm m}$) 
in the right-handed frame with $\hat z$-axis along the dipole
axis) 
and $\theta \ll 1$~rad is the magnetic colatitude of a footpoint of the parent magnetic field line
at the neutron star
surface. 
For field lines originating at the polar-cap rim one should take then
$\theta = \theta_{\rm pc}$, 
where $\theta_{\rm pc} = \arcsin(\sqrt(R_{\rm NS}/R_{\rm lc})) \approx 1.45\cdot10^{-2}/\sqrt{P}$ radians,
and $R_{\rm lc} = cP/2\pi$ is the light cylinder.
Note, that the escape energy of eq.(\ref{eesc0}) is symmetrical with respect to the dipole axis
and goes to infinity at the magnetic pole ($\theta = 0$).
Such a behaviour is not the case, however, when the rotational effects discussed in section~3
are taken into account.

\begin{figure}
\centering
\resizebox{\hsize}{!}{\includegraphics{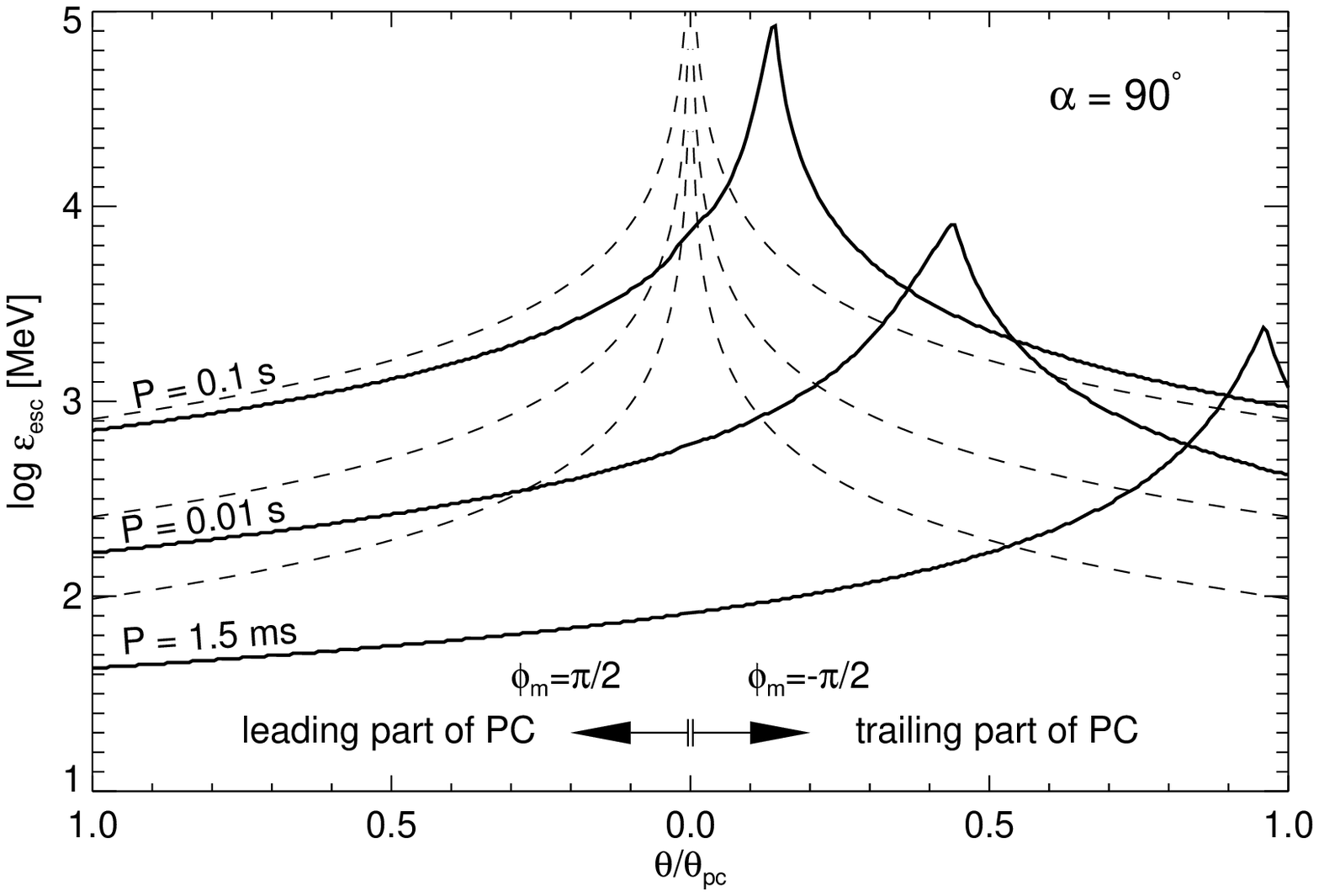}}
\caption{The escape energy $\varepsilon_{\rm esc}$ of photons from the polar cap surface of an orthogonal rotator 
with $B_{\rm pc} = 10^{12}$G
is shown as a function of normalized magnetic colatitude $\theta/\theta_{\rm pc}$ of the emission points.
The points are assumed to lay along the cross-section of the polar cap surface
with the equatorial plane of rotation, thus location of each point is determined by $\theta/\theta_{\rm pc}$
in the range $[0,\,1]$, and the magnetic azimuth $\phi_{\rm m}$  equal either to
$\pi/2$ (for the leading half of the polar cap) or $-\pi/2$ (for the trailing half).
Three solid lines
are labelled with the corresponding spin periods $P$ of $0.1$~s, $10$~ms, and $1.5$~ms.
Moreover, to illustrate the difference with 
the case when rotational effects are ignored,
each solid line is accompanied by a dashed line drawn
according to eq.(\ref{eesc0}).
}
\end{figure}

In Figs.~7, 8, and 9  we present the values of
$\varepsilon_{\rm esc}$ obtained numerically
for fast  rotators 
with $B_{\rm pc} = 1$ TG, and emission points located on the polar cap surface,
i.e. $r_{\rm em} = R_{\rm NS}$ was assumed everywhere.
Let us begin with the case of orthogonal rotators (i.e. $\alpha = 90^\circ$) - these are
shown in Figs.~7 and 8. Here we consider emission points lying
along the crossection of polar cap surface
with the equatorial plane of rotation (hence, the results would be relevant for
observers located at $\zeta_{\rm obs} = 90^\circ$).
Fig.~7
shows how $\varepsilon_{\rm esc}$ varies with location of the emission point across the polar cap.
The location of each point is determined by the normalized magnetic colatitude $\theta/\theta_{\rm pc}$
in the range $[0,\,1]$ and the magnetic azimuth $\phi_{\rm m}$  equal either to
$\pi/2$ (for the leading half of the polar cap)
or 
$-\pi/2$ (for the trailing half).
The spin periods of $0.1$~s, $10$~ms, and $1.5$~ms were considered. 

In order to quantify the asymmetry in $\varepsilon_{\rm esc}$ between the leading and the trailing parts
of the polar cap we introduce the following parameter:
\begin{equation}
{\cal R}_{\rm esc}(\theta) = \frac{\varepsilon_{\rm esc}^{\rm tp}}
{\varepsilon_{\rm esc}^{\rm lp}},
\label{resc}
\end{equation} 
where 
\begin{eqnarray}
\lefteqn{\varepsilon_{\rm esc}^{\rm lp} \equiv 
\varepsilon_{\rm esc}(\phi_{\rm m}=\pi/2,\ \theta),}\nonumber\\
\lefteqn{\varepsilon_{\rm esc}^{\rm tp} \equiv 
\varepsilon_{\rm esc}(\phi_{\rm m}=-\pi/2,\ \theta).}
\label{resc1}
\end{eqnarray} 
For example, ${\cal R}_{\rm esc}(\theta_{\rm pc})$
gives the asymmetry between two opposite points located on the outer rim of the polar cap.
For spin periods around $0.1$~s, typical for the known gamma-ray pulsars,
${\cal R}_{\rm esc}(\theta_{\rm pc})$ remains
close to unity; e.g. for the case of $P = 0.1$~s in Fig.7, ${\cal R}_{\rm esc}(\theta_{\rm pc}) 
\simeq 1.3$. However, for $P$ smaller than $0.01$~s the parameter 
${\cal R}_{\rm esc}(\theta_{\rm pc})$ becomes definitely larger:
$\sim 2.5$ for $P = 0.01$~s and $\sim 30$ for $P = 1.5$~ms. 
Thus, for pulsars with $P \la 0.01$~s and large inclination angles $\alpha$
we predict a notable difference (in excess of half a decade in photon energy)
between the positions of high-energy spectral cutoff
in the gamma-ray emission from the leading and the trailing part of the
polar cap.

Another interesting implication of fast rotation is that 
$\varepsilon_{\rm esc}$ has 
finite values for any $\theta$, including $\theta = 0$ 
(the magnetic pole), in contrast to the `static' case of eq.~(\ref{eesc0}).
This can understood in the following way: 
consider 
emission points with decreasing 
colatitude $\theta$ in the leading part of the polar cap. 
As  we approach the dipole axis ($\theta \rightarrow 0$), 
$\varepsilon_{\rm esc}$ increases because
the decreasing curvature of magnetic field lines leads to smaller angles
between $\vec B^\prime$ and the photon propagation direction $\hat \eta^\prime$
in the corotating frame CF. However, photon 
trajectory bends backwards in the CF  (see dashed lines in Fig.~1), 
which implies that
also photons emitted at $\theta = 0$
along the straight dipolar axis will
quickly encounter $B^\prime_\perp \neq 0$, thus being subject to the magnetic absorption.
Entering now the trailing part of the polar cap leads to
further increase 
of $\varepsilon_{\rm esc}$. This is
because magnetic field lines start to bend in the same direction
as the photon trajectory in the CF 
(in other words - the efficiency
of absorption {\em decreases} for emission points in the trailing
part of the polar cap). Eventually, at some point (we denote it as $\theta_0$) the escape energy
reaches a maximum. This is the point where 
the magnetic field slippage along with the aberration of photon
direction ensure small angles between 
$\vec B^\prime$ and $\hat \eta^\prime$ over 
large distances in the photon trajectory. Therefore,
the faster is the rotation,
the larger is the colatitude $\theta_0$ of that point. 
For example, $\theta_0/\theta_{\rm pc} \simeq 0.15$
for $P = 0.1$~s whereas 
$\theta_0/\theta_{\rm pc} \simeq 0.42$ for $P = 0.01$~s (see Fig.~7).
For $P=1.5$ ms this maximum occurs~\footnote{It is easy to reproduce the behaviour of $\theta_0/\theta_{\rm pc}$
as a function of $P$ in Fig.~7 with analytic formula: 
The maximal value of
$B_{\perp}(r)/B_{\rm pc}$
encountered in a static dipolar field by a photon emitted along the local field line at $\theta_0$
is approximately equal to $0.1\, \theta_0$ (Sturrock \cite{sturrock}, also Fig.1 
in Rudak \& Ritter \cite{rr94}). This occurs always
at $r_0 = 4/3\, R_{\rm NS}$,
regardless the value of $\theta_0$. The angle $\psi$ between the photon propagation direction and
the local field line at $r_0$ is
$\psi_0 
\approx B_{\perp}(r_0)/ B(r_0)$ 
and therefore $\psi_0 \approx 0.1\, \theta_0  (4/3)^3$.
To minimize $B_{\perp}^\prime (r_0)$ as much as possible in the case of rotation,
the aberration angle due to local linear velocity $\beta(r_0) = 4/3\,R_{\rm NS}/R_{\rm lc}$ should be
close to
$\psi_0$. Since $\beta(r_0)\ll 1$, the aberration angle is $\sim \beta(r_0)$.
Therefore, we obtain the condition 
$0.1\,(4/3)^2\,\theta_0 \approx \theta_{\rm pc}^2$ which
gives
\begin{equation}
\frac{\theta_0}{\theta_{\rm pc}} \approx 0.08\, P^{-1/2}.
\label{pmax}
\end{equation}
This formula overestimates the locations of maxima in Fig.~7
by a factor $\sim 2$ only.} 
close to the outer rim (in the
trailing part) and therefore huge asymmetry 
with respect to the leading rim is predicted: 
${\cal R}_{\rm esc}(\theta_{\rm pc}) \simeq 30$.
With further increase of $\theta$ 
(towards the trailing rim of the polar cap)
local magnetic field lines
start to bend stronger than the photon trajectory in the CF, 
and this is why
$\varepsilon_{\rm esc}$ should now decrease.
However, this decrease never compensates the asymmetry 
in $\varepsilon_{\rm esc}$ at $\theta_{\rm pc}$
with respect to $\theta = 0$,
i.e. one always ends up with  ${\cal R}_{\rm esc}(\theta_{\rm pc}) > 1$.

\begin{figure}
\centering
\resizebox{\hsize}{!}{\includegraphics{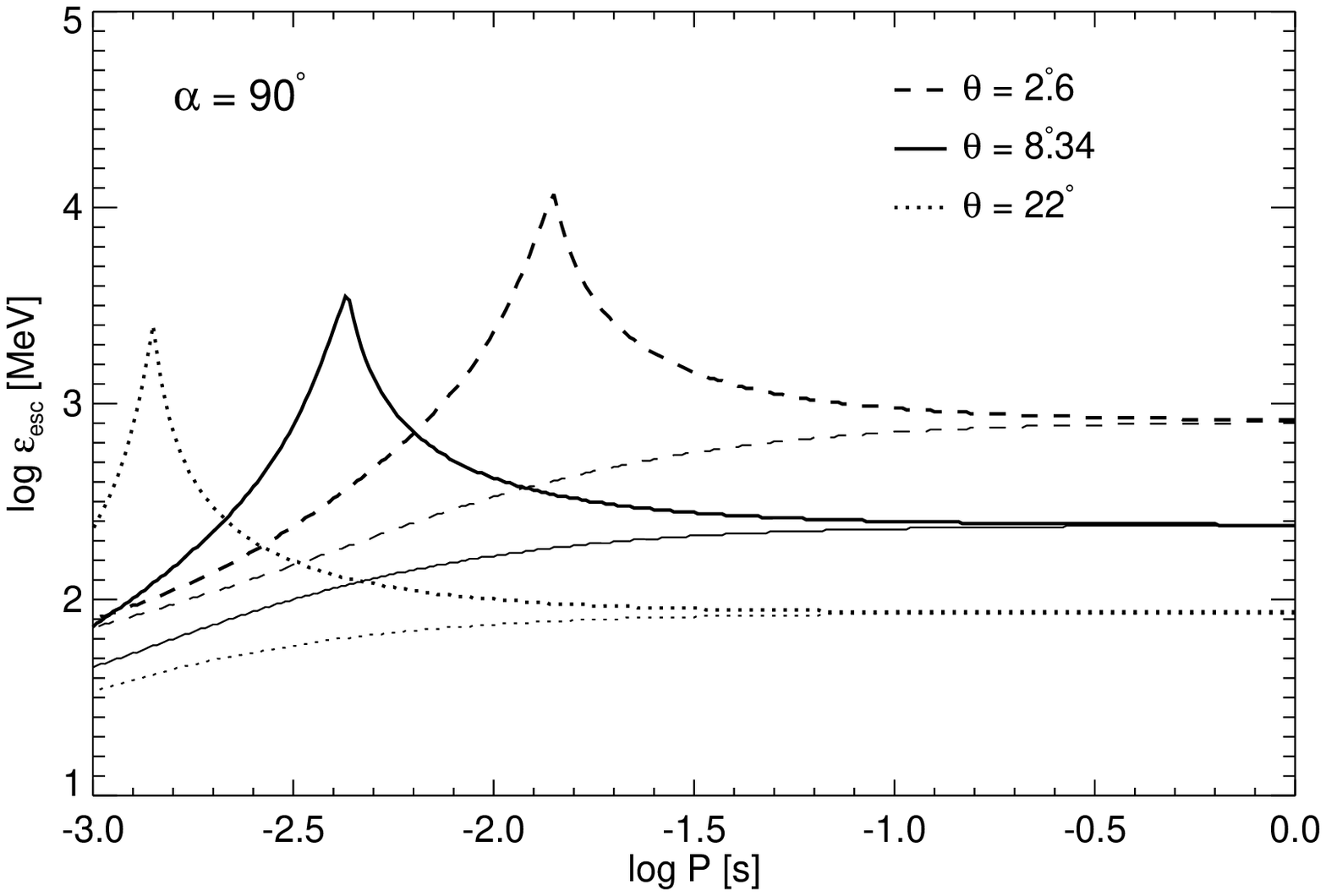}}
\caption{Escape energy at fixed magnetic colatitude $\theta$ 
as a function of rotation period. The dashed, solid, and dotted lines
correspond to $\theta=2.6^\circ$, $8.3^\circ$, and $22^\circ$,
respectively. The thin lines correspond to points on the leading side of the
dipole axis (with spherical coordinates ($R_{\rm NS}$, $\pi/2$,
$\theta$)).
The thick lines correspond to points on the trailing side
(with ($R_{\rm NS}$, $-\pi/2$, $\theta$)).
The curves are for emission in the equatorial plane of orthogonal rotator.
}
\label{pdep}
\end{figure}

Figure~8 presents escape energy $\varepsilon_{\rm esc}^{\rm lp}$ and 
$\varepsilon_{\rm esc}^{\rm tp}$
in function of spin period $P$.
This energy was calculated for a fixed position $\theta_{\rm fxd}$ of emission point,
in order to highlight its dependence
on rotation.
We chose three pairs of oppositely located emission points at: 
$\theta _{\rm fxd} = 2^\circ.6$, $8^\circ.3$, and $22^\circ$, 
which corespond to $\theta_{\rm pc}$ for $P = 0.1$ s, $10$
ms, and $1.5$ ms, respectively. As in Fig.~7,
the emission points were placed at the neutron star surface,
in the equatorial plane of orthogonal rotator.
In the case of slow rotation ($P \sim 1$ s), the values of $\varepsilon_{\rm
esc}^{\rm lp}(\theta_{\rm fxd})$ for the leading point and 
$\varepsilon_{\rm esc}^{\rm tp}(\theta_{\rm fxd})$ for the trailing point 
are practically identical, 
and well approximated by eq.(\ref{eesc0}).
As rotation becomes faster ($P$ around $\sim 0.1$ s) 
$\varepsilon_{\rm esc}^{\rm lp}(\theta_{\rm fxd})$ and
$\varepsilon_{\rm esc}^{\rm tp}(\theta_{\rm fxd})$ 
start to diverge due to the asymmetric influence of $\vec E$.
At even shorter  periods, below $\sim 0.03$ s,
the maxima in $\varepsilon_{\rm esc}^{\rm tp}$ are reached 
(at the values of $P$, which can be reproduced by solving eq.(\ref{pmax})
with $\theta_{\rm fxd}$ in place of $\theta_0$)
because slippage of field lines starts to be important (see previous paragraph).
For increasing $\theta _{\rm fxd}$ ($2^\circ.6$, $8^\circ.3$, and $22^\circ$ in Fig.~8),
the asymmetry parameter ${\cal R}_{\rm esc}(\theta _{\rm fxd})$ decreases 
for slow rotators ($P \ga 0.1$~s), and it increases for fast (millisecond) rotators.

\begin{figure}
\centering
\resizebox{\hsize}{!}{\includegraphics{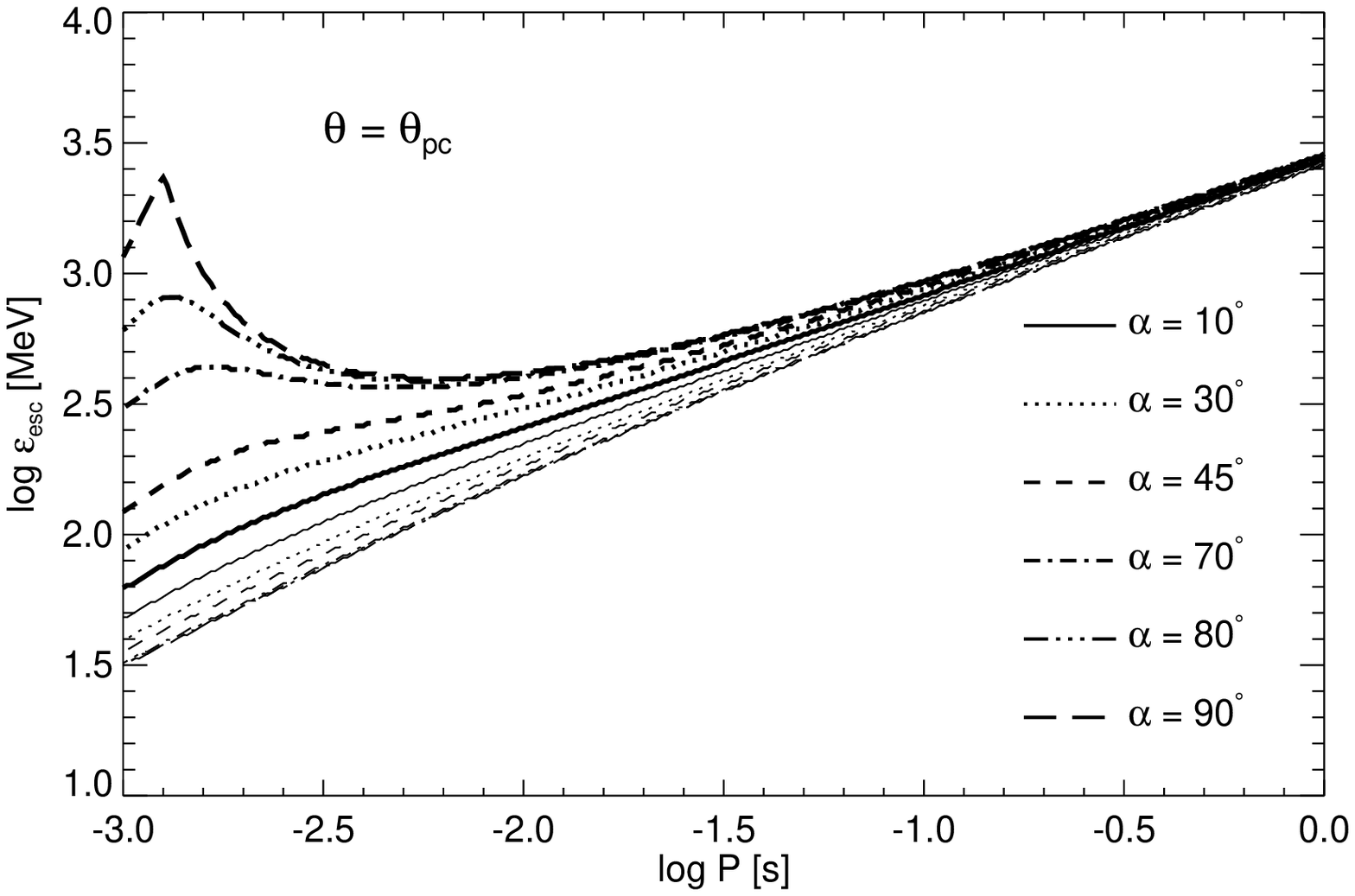}}
\caption{Escape energy 
for the leading and the trailing emission points
at the polar cap rim as a function of rotation
period.
Six uppermost lines (thick) correspond to the trailing points.
Thin lines are for the leading points
(the lines for $\alpha = 70^\circ$, $80^\circ$, and $90^\circ$ overlap).
}
\label{tiltdep}
\end{figure}

In Fig.~9 we present $\varepsilon_{\rm esc}^{\rm lp}(\theta_{\rm pc})$ 
and $\varepsilon_{\rm esc}^{\rm tp}(\theta_{\rm pc})$  as a function of spin period $P$
for a set of dipole inclinations $\alpha$.
Note, that unlike in Fig.~8, the emission points are now placed at $\theta =
\theta_{\rm pc}$, i.e.~at the rim of polar cap corresponding to $P$. 
In the case of  small inclinations 
(eg.~$\alpha = 10^\circ$),
the resulting ratio ${\cal R}_{\rm esc}(\theta_{\rm pc})$ remains close to unity 
even for millisecond periods.
The general increase in $\varepsilon_{\rm esc}$ with $P$ increasing, 
noticeable in Fig.~9, reflects
the approach of emission points to the dipole axis,
where the  curvature of magnetic field lines is small.
The trend is well described by $\varepsilon_{\rm esc}\propto \sqrt{P}$ 
as given by eq.~(\ref{eesc}).
In the range of spin periods below $\sim 0.1$~s and with 
$\alpha \ga 45^\circ$, the difference between $\varepsilon_{\rm esc}^{\rm
lp}$ and $\varepsilon_{\rm esc}^{\rm tp}$ becomes pronounced, especially
for highly inclined millisecond pulsars.
If detected by GLAST, high-energy emission
from such objects would provide an ideal test of the polar cap model.

However, for the asymmetry effects to be detectable, 
an additional condition (apart from short period
and large inclination) must be fulfilled: the viewing
geometry must be of the `on-beam' type, i.e.~the observer's line of sight 
must cross
the narrow beam of radiation at the high-energy cutoff - where 
the magnetic absorption operates.
If this is the case, 
the asymmetry in the absorption
may be noticeable even in the phase-averaged spectra.
As an example, we show in Fig.~10 the phase-averaged spectrum 
calculated for a millisecond pulsar with
$P=2.3$ ms, $B_{\rm pc} = 10^9$ G, $\alpha = 60^\circ$, and for $\zeta_{\rm
obs} = 60^\circ$. For this rotator we obtain 
$\varepsilon_{\rm esc}^{\rm lp} \simeq 10^5$ MeV 
and $\varepsilon_{\rm esc}^{\rm tp} \simeq 5\cdot 10^5$ MeV 
at the rim of the polar cap.
As a result of this rotationally induced asymmetry in magnetic absorption
for the leading and the trailing peak
the spectrum at its  high-energy cutoff assumes a step-like shape:
below $\simeq 10^5$ MeV the spectrum consists of photons from 
both the leading
and the trailing peak, whereas between $\sim 10^5$ MeV 
and $\sim 5\cdot 10^5$ MeV
only photons of the trailing peak contribute to the spectrum;
at $\varepsilon \simeq \varepsilon_{\rm esc}^{\rm lp}$ the level of
the spectrum drops by a factor of $\sim 2$.
In these particular calculations of the spectrum, we assumed that 
the density distribution of primary electrons over the polar cap 
is dominated by an outer-rim component (see the captions to Fig.~10 for details of the distribution).

\begin{figure}
\centering
\resizebox{0.6\textwidth}{!}{\includegraphics{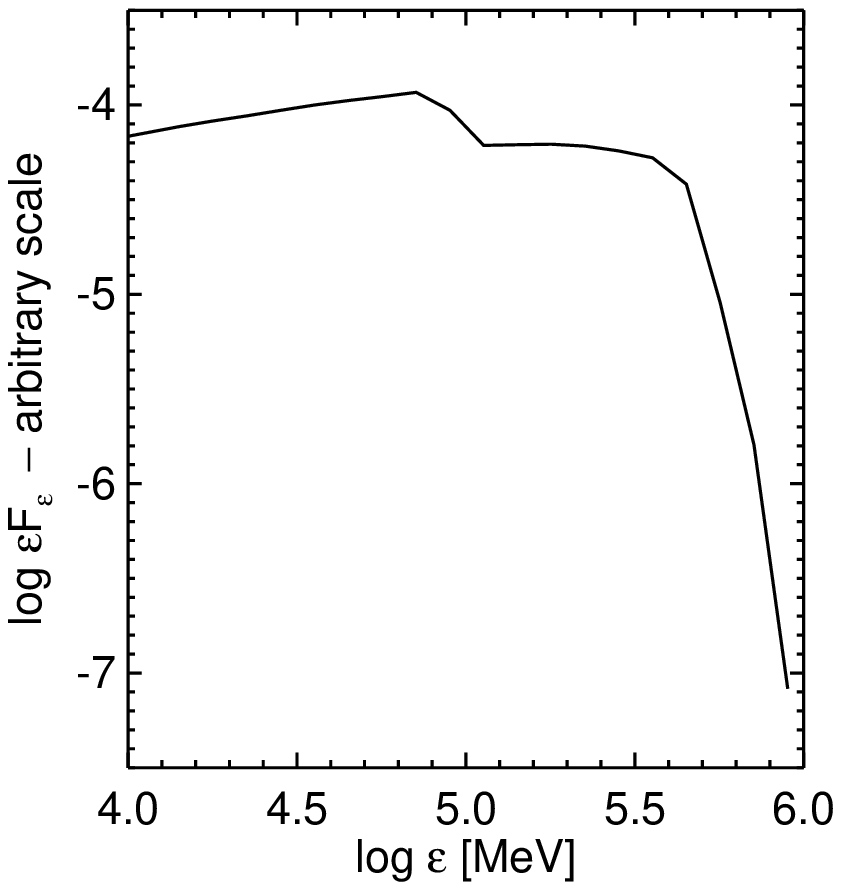}}
\caption{Theoretical "on-beam" spectrum of high-energy emission from 
a millisecond pulsar with $P = 2.3$ ms, $B_{\rm pc} = 10^9$ G,
and 
$\zeta_{\rm obs} = \alpha = 60^\circ$. 
The density distribution of primary
electrons
over the polar cap, which was assumed in this calculation, 
had a maximum at the polar cap rim and was uniform
in magnetic azimuth $\phi_{\rm m}$.
The magnetic colatitude profile of this distribution
had a gaussian shape centered at 
$\theta = \theta_{\rm pc}$, with
a width $\sigma = 0.05\ \theta_{\rm pc}$. The initial energy of primary
electrons was equal to $2\cdot 10^7$ MeV.
Note the step-like decline near $\sim 100$ GeV.
}
\label{steplike}
\end{figure}

If, however, an emission from the inner part 
of the polar cap were to contribute considerably to the outer-rim emission, the step-like
shape shown in Fig.~10 would be smoothed out, because of contribution
of many spectra with different values of $\varepsilon_{\rm esc}$.
In such a case 
the polar cap origin of the observed radiation
could be easily revealed by noting
strong differences between the high-energy spectral cutoffs
in different ranges of the rotational phase (i.e. phase-resolved spectra would have
to be obtained).
The first obvious candidate to check for this effect (e.g. with GLAST) seems to be
J0218$+$4232 -- the only gamma-ray pulsar among all millisecond pulsars
(Kuiper et al.~\cite{k2000}). However, this pulsar appears to be 
a candidate for an `off-beam' case (see Dyks \& Rudak \cite{dr2002} for details).

The strength of magnetic field $B_{\rm pc}$ practically does not affect 
the shapes of curves shown in Figs.~7 -- 9. 
The magnetic field only acts as a scaling factor: $\varepsilon_{\rm esc} 
\propto B^{-1}$; cf.~eq.~(\ref{eesc}).

\section{Discussion}

We have shown that pulsar rotation induces an asymmetry in the 
magnetic absorption rate with respect to the magnetic dipole axis.
Its consequences are potentially interesting in constraining
the phase-space of parameters in the polar cap models of high-energy radiation,
provided that very high quality gamma-ray
data (e.g. as expected from GLAST) are at hand.
Its magnitude depends mainly on the linear velocity $\beta$ of the magnetosphere at  
sites of particle acceleration and magnetic photon absorption.
When the region of electron acceleration is placed just above the  neutron star surface
rotation does not produce any detectable effects even for relatively
fast rotating young gamma-ray pulsars.
However, it has been argued that at least in the case of the Vela pulsar, such a situation 
is difficult to reconcile with the spectral high-energy cutoff at about 10~GeV (e.g. Dyks et al.~\cite{alic}).
We find then that raising the accelerator up to $\sim 4$ neutron star radii
(in the spirit of Harding \& Muslimov \cite{hm98}) 
above its polar cap produces asymmetric gamma-ray pulse profiles even in the case
of nearly aligned rotators with a spin period of $P\sim 0.1$~s.
The resulting features - softer spectrum of the leading peak and the
dominance of the trailing peak above $\sim 5$ GeV - do
agree qualitatively with the EGRET data of the bright gamma-ray pulsars (Thompson~\cite{thompson2001}). 

We are far from concluding that the rotation effects alone
can account for the observed asymmetry  
in the double peaks of the bright EGRET pulsars. 
On the contrary - some axial asymmetry intrinsic to the region of electron acceleration is 
inevitable in order to explain the double-peak properties at $\> 100$~MeV
of Geminga and B1706-44, where the leading peak is weaker than the trailing peak.
Strong deviations of the actual magnetic field structure
from the pure dipole at the stellar surface 
(eg.~Gil et al.~\cite{gmm2002})
might be responsible for maintaining
axial asymmetry at the site of electron acceleration (unlike the symmetric initial conditions introduced in Section 2).
This in turn would lead to electromagnetic cascades whose properties vary with magnetic azimuth.
It is important, however, that the propagation effects due to rotation
work in the right direction, i.e.~they explain qualitatively
the observed weakening of the leading peak with respect to the trailing peak.   
We emphasize that this weakening 
occurs only in the vicinity of the (phase-averaged) high-energy spectral cutoff,
where the flux level decreases significantly.

Another consequence of the magnetic absorption of high energy photons
is a noticeable change in the separation $\Delta^{\rm peak}$ between the two peaks in the pulse,
taking place near the high-energy spectral cutoff (Dyks \& Rudak \cite{dr2000}).
In the  model discussed above, with electrons ejected 
only from a rim of the polar cap, the higher energy of photons requires 
higher emission altitudes to avoid absorption.
Therefore, a slight increase in $\Delta^{\rm peak}$ is visible 
in the three lowermost pulse profiles in Fig.~\ref{profiles}\thinspace b.
However, if the emission from the interior of the polar cap were included,
just the opposite behaviour would occur: $\Delta^{\rm peak}$ would decrease
near the high-energy cutoff in the spectrum.
This is because in this case of a ``filled polar cap tube",
the highest energy
non-absorbed photons are emitted closer to the magnetic dipole
axis (see Fig.~2 in Dyks \& Rudak \cite{dr2000}).
The latter case agrees qualitatively with the marginal decrease in peak
separation found in the EGRET data for Vela  
(Kanbach \cite{kanbach99}).

Stimulated by high-quality observations of
gamma-ray pulsars anticipated with GLAST we analysed in Sect.~5 the importance
of rotation-driven asymmetry in magnetic absorption for a broad range of pulsar parameters.
A decline in gamma-ray flux at high-energy spectral cutoff
should inevitably be accompanied by strong changes in pulse profiles: whereas at lower
photon energies the profile is determined by the density
distribution of primary electrons over the polar cap and the efficiency 
of photon emission mechanism, in the vicinity of the cutoff
it becomes additionally constrained by likely high values of the asymmetry parameter 
${\cal R}_{\rm esc}(\theta_{\rm pc})$ - the situation anticipated for
fastly rotating  ($P < 0.01$~s), and highly inclined ($\alpha \ga 45^\circ$) 
pulsars.

\begin{acknowledgements}
 We thank V.S.~Beskin and A.K.~Harding for useful comments 
 on the issue of magnetospheric distortions. We are grateful to Gottfried~Kanbach
 for providing us with the EGRET data on Vela, and to Aga Wo{\' z}na 
 for calculating the P2/P1 ratios used in Fig.\ref{p1p2}. 
 We acknowledge comments and stimulating suggestions made by the anonymous referee.
 JD appreciates Young Researcher Scholarship of Foundation
 for Polish Science.
 This work was supported by KBN (grants 2P03D02117 and 5P03D02420) and NCU 
 (grant 405A).  
\end{acknowledgements}

\end{document}